\documentclass[twocolumn,twoside,slac_two]{revtex4-1}
\usepackage{graphicx}
\usepackage{fancyhdr}
\usepackage{hyperref}
\hypersetup{
    colorlinks=true,
    urlcolor=magenta,
}
\begin{document}

\title{Precision measurement of antiproton to proton ratio with the Alpha Magnetic Spectrometer on the International Space Station}

\author{F.~Nozzoli on behalf of the AMS Collaboration}
\affiliation{INFN, Sezione di Perugia, I-06100 Perugia, ASI Science Data Center, via del Politecnico s.n.c., I-00133 Roma Italy}

\begin{abstract}
  A precision measurement by AMS of the antiproton-to-proton flux ratio in
  primary cosmic rays in the absolute rigidity range from 1 to 450 GV is presented based on $3.49\times10^5$
  antiproton events and $2.42\times10^9$ proton events.
  Above $\sim60$ GV the antiproton to proton flux
  ratio is consistent with being rigidity independent. A decreasing behaviour is expected for this ratio considering the traditional models for the secondary antiproton flux. 
\end{abstract}

\maketitle

\thispagestyle{fancy}


The measurement of the antiproton-to-proton flux ratio in primary Cosmic
Rays (CR) is reported in the absolute rigidity range 1-450 GV. 
This measurement is based on 3.49 $\times 10^5$ antiproton events
and 2.42 $\times 10^9$ proton events collected by the Alpha Magnetic Spectrometer,
AMS
\cite{ref:AMS2_first,ref:AMS2_fraction,ref:AMS2_epfluxes,ref:AMS2_allele,ref:AMS2_protons,ref:AMS2_He,ref:AMS2_pbar,ref:AMS2_BC},
on the International Space Station, ISS, from May
19, 2011 to May 26, 2015.
The experimental data on antiprotons
are limited \cite{ref:BESS,ref:PAMELA} because of
their very low flux intensity, up to this measurement only
a few$\times 10^3$ antiprotons were observed in the cosmic radiation.
In the measurement of the antiproton component of the cosmic radiation a very large background is expected from the most abundant proton one:  
for each antiproton there are
approximately $10^4$ protons, therefore,
to measure the antiproton flux to 1$\%$ accuracy requires a
separation power of $\sim 10^6$. The sensitivity of antiprotons to
exotic CR sources, as dark matter annihilations, as well as to new
phenomena in the propagation of CR in the galaxy
is complementary to the sensitivity
of the measurements of CR positrons. In particular, AMS has accurately measured the excess in the
positron fraction to 500 GeV \cite{ref:AMS2_first,ref:AMS2_fraction} and
this data generated many interesting theoretical models
including collisions of dark matter particles, astrophysical sources,
and collisions of CR (see e.g.
\cite{ref:tom_last,ref:tom_bay,ref:tom_PHe,ref:tom_unc,ref:tom_2hal,ref:tom_cfE,ref:tom_posfra,ref:donato}).
Some of these models also
include specific predictions for the antiproton flux and the
antiproton-to-proton flux ratio in CR.

\noindent {\bf Detector.} The description of the AMS detector is
presented in \cite{ref:AMS2_first,ref:AMS2_fraction,ref:AMS2_epfluxes,ref:AMS2_allele,ref:AMS2_protons,ref:AMS2_He,ref:AMS2_pbar,ref:AMS2_BC}.
All detector elements are used for particle identification in the present analysis:
the silicon tracker TRK, the permanent magnet, the time of flight
counters TOF, the anticoincidence counters ACC,
the transition radiation detector TRD, the ring imaging
Cherenkov detector RICH, and the electromagnetic
calorimeter ECAL.
\begin{figure}[t]
\centering
\includegraphics[width=64mm]{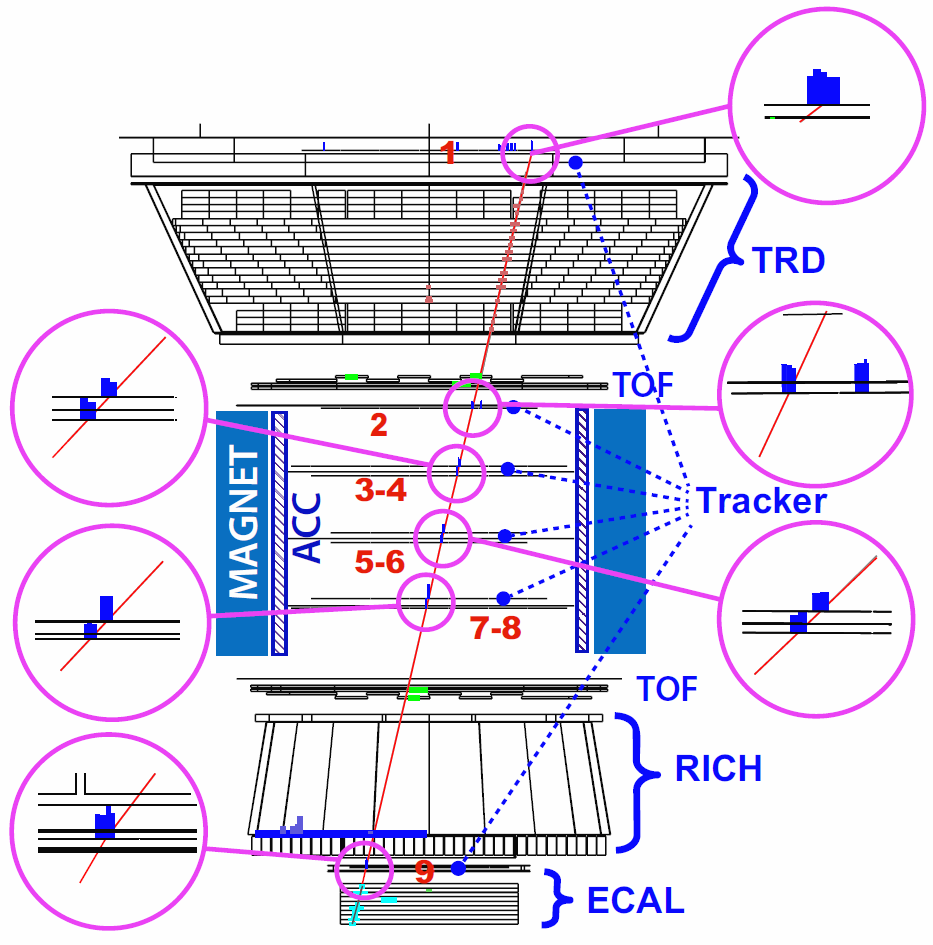}
\caption{
  An event display in the bending plane of an antiproton. The red line indicates
  the reconstructed trajectory or track. The insets indicate the matching of the track to the pulse
  heights measured in each layer of the tracker. This downward-going $\vert Z \vert$ = 1 event is identified as
  an antiproton with R = -435 GV, $\Lambda_{TRD}$ = 0.908, and $\Lambda_{CC}$ = 0.983. See \cite{ref:AMS2_pbar} for details.
} \label{detector}
\end{figure}
The tracker, with its nine layers, is used to
measure the rigidity R (momentum per unit of charge)
of cosmic rays and to differentiate between positive and
negative particles. 
The first layer (L1)
is at the top of the detector, the second (L2) just above the
magnet, six (L3 to L8) within the bore of the magnet, and
the last (L9) just above the ECAL. L2 to L8 constitute the
inner tracker. For $\vert Z \vert$ = 1 particles the maximum detectable
rigidity, MDR, is 2 TV and the charge resolution is
$\Delta$Z = 0.05. The TOF measures $\vert Z \vert$ and velocity with a
resolution of $\Delta \beta / \beta^2  = 4\%$. The ACC has 0.99999 efficiency
to reject particles entering the inner tracker from
the side.
The TRD separates $\bar{p}$ and p from $e^-$ and $e^+$ using the
$\Lambda_{TRD}$ estimator constructed from the ratio of the log-likelihood
probability of the $e^{\pm}$ hypothesis to that of the $\bar{p}$ or p
hypothesis in each layer. Antiprotons and protons,
which have $\Lambda_{TRD}$ $\sim 1$, are efficiently separated from electrons
and positrons, which have $\Lambda_{TRD}$ $\sim$ 0.5. The RICH has a
velocity resolution $\Delta \beta / \beta = 0.1\%$ for $\vert Z \vert$ = 1 to ensure
separation of $\bar{p}$ and p from light particles ($e^{\pm}$ and $\pi^{\pm}$)
below 10 GV. The ECAL is used to separate $\bar{p}$ and p from $e^-$ and $e^+$
when the event can be measured by the ECAL.
Antiprotons are separated from charge confused protons,
that is, protons which are reconstructed in the tracker
with negative rigidity due to the finite tracker resolution or
due to interactions with the detector materials, by means of a charge
confusion estimator $\Lambda_{CC}$ defined with a boosted
decision tree technique \cite{BDT}. The estimator combines
information from the tracker such as the track $\chi^2$/d.o.f.,
rigidities reconstructed with different combinations of
tracker layers, the number of hits in the vicinity of the
track, and the charge measurements in the TOF and the
tracker. With this method, antiprotons, which have
$\Lambda_{CC}\sim$+1, are efficiently separated from charge confused
protons, which have $\Lambda_{CC}\sim$-1.
An example of a 435 GV antiproton crossing the AMS sub-detectors
is given in Fig. \ref{detector}.

\noindent {\bf Event selection and data samples.}
Over 65 billion
CR events have been recorded in the first 48 months
of AMS operations. Only events collected during normal
detector operating conditions are used in this analysis. This
includes the time periods when the AMS z axis is pointing
within 40$^o$ of the local zenith and when the ISS is not in the
South Atlantic Anomaly. Data analysis is performed in 57
absolute rigidity bins. The same binning as in our proton
flux measurement \cite{ref:AMS2_protons}
was chosen below 80.5 GV. Above
80.5 GV two to four proton bins are combined to
ensure sufficient antiproton statistics.
Events are selected requiring a track in the TRD and in
the inner tracker and a measured velocity $\beta >$ 0.3 in the
TOF corresponding to a downward-going particle. To
maximize the number of selected events while maintaining
an accurate rigidity measurement, the acceptance is
increased by releasing the requirements on the external
tracker layers, L1 and L9. Below 38.9 GV neither L1 nor
L9 is required. From 38.9 to 147 GV either L1 or L9 is
required. From 147 to 175 GV only L9 is required. Above
175 GV both L1 and L9 are required. In order to maximize
the accuracy of the track reconstruction, the  $\chi^2$/d.o.f. of the
reconstructed track fit is required to be less than 10 both in
the bending and nonbending projections. The dE/dx
measurements in the TRD, the TOF, and the inner tracker
must be consistent with $\vert Z \vert$ = 1. To select only primary
CR, the measured rigidity is required to exceed the
maximum geomagnetic cutoff by a factor of 1.2 for either
positive or negative particles within the AMS field of view.
The cutoff is calculated by backtracing using the
most recent IGRF geomagnetic model.
Events satisfying the selection criteria are classified into
two categories: positive and negative rigidity events. A
total of 2.42 $\times 10^9$ events with positive rigidity are selected
as protons. They are 99.9$\%$ pure protons with almost no
background. Deuterons are not distinguished from protons,
their contribution decreases with rigidity: at 1 GV it is less
than 2$\%$ and at 20 GV it is 0.6$\%$. The effective
acceptance of this selection for protons is larger than in our
proton flux publication \cite{ref:AMS2_protons}.
This is because there is no
strict requirement that selected particles pass through the
tracker layers L1 and L9 (see above) leading to a much
larger field of view at low rigidities and, therefore, to a
significant increase in the number of protons.
The negative rigidity event category comprises both
antiprotons and several background sources: electrons,
light negative mesons ($\pi^-$ and a negligible amount of
$K^-$) produced in the interactions of primary CR
with the detector materials, and charge confused protons.
The contributions of the different background sources vary
with rigidity. For example, light negative mesons are
present only at rigidities below 10 GV, whereas charge
confusion becomes noticeable only at high rigidities.
Electron background is present at all rigidities. The
combination of information from the TRD, TOF, tracker,
RICH, and ECAL enables the efficient separation of the
antiproton signal events from these background sources
using a template fitting technique. The number of observed
antiproton signal events and its statistical error in the
negative rigidity sample are determined in each bin by
fitting signal and background templates to data by varying
their normalization. As discussed below, the template
variables used in the fit are constructed using information
from the TOF, tracker, and TRD. The distribution of the
variables for the template definition is the same for
antiprotons and protons if they are both reconstructed with
a correct charge-sign. This similarity has been verified with
the Monte Carlo simulation and the antiproton and
proton data of 2.97 $\leq \vert R \vert <$ 18.0 GV. Therefore, the signal
template is always defined using the high-statistics proton
data sample. Three overlapping rigidity regions with different
types of template function are defined to maximize the
accuracy of the analysis: low absolute rigidity region (1.00-4.02 GV),
intermediate region (2.97-18.0 GV), and high
absolute rigidity region (16.6-450 GV). In the overlapping
rigidity bins, the results with the smallest error are selected.
At low rigidities, a cut on the TRD estimator $\Lambda_{TRD}$ and
the velocity measurement in the TOF are important to
differentiate antiprotons from light particles ($e^-$ and $\pi^-$).
Therefore, the mass distribution, calculated from the
rigidity measurement in the inner tracker and the velocity
measured by the TOF, is used to construct the templates and
to differentiate between the antiproton signal and the
background. The background $e^-$ and $\pi^-$ templates are
defined from the data sample selected using information
from the TRD, the RICH, and also the ECAL, when the
event can be measured by the ECAL.
\begin{figure}[t]
\centering
\includegraphics[width=64mm]{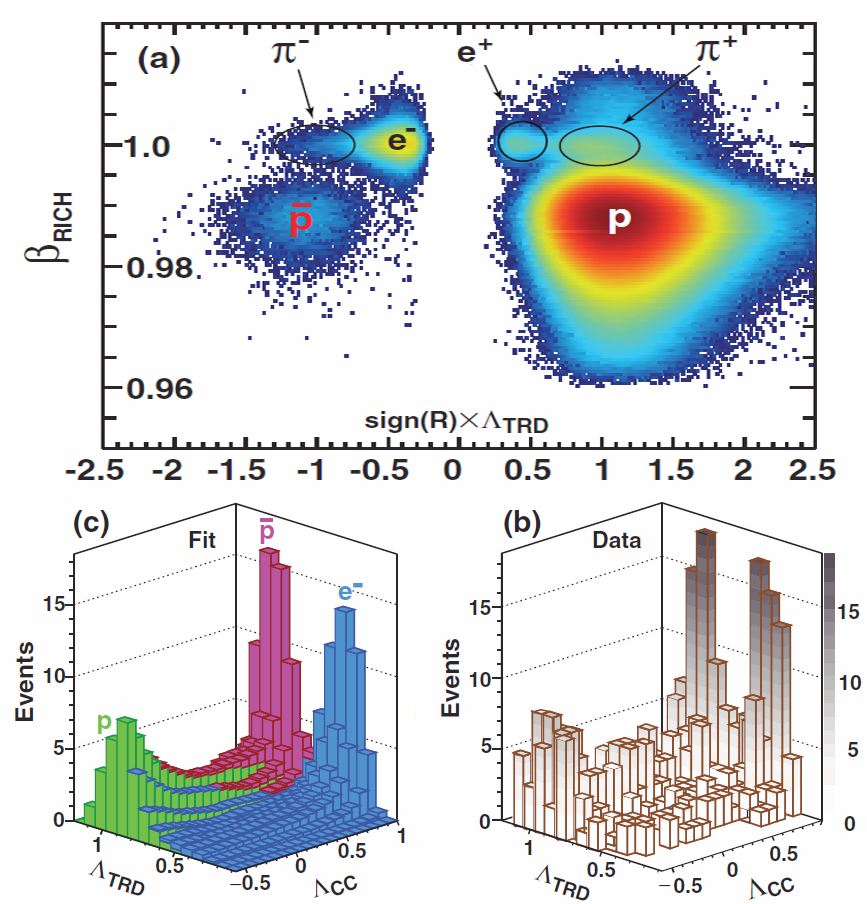}
\caption{(a) Negative rigidity and positive rigidity data samples
  in the [$\beta_{RICH}$ - sign(R)$\times \Lambda_{TRD}$] plane for the absolute rigidity
  range 5.4-6.5 GV. The contributions of $\bar{p}$, p, $e^+$, $e^-$, $\pi^+$, and $\pi^-$
  are clearly seen. The antiproton signal is well separated from the
  backgrounds. (b) For negative rigidity events, the distribution of
  data events in the ($\Lambda_{TRD}$ - $\Lambda_{CC}$) plane for the absolute rigidity bin
  175-211 GV. (c) Fit with $\chi^2$/d.o.f. = 138/154 of the antiproton
  signal template (magenta), the electron background template
  (blue), and the charge confused proton background template
(green) to the data in (b). See \cite{ref:AMS2_pbar} for details.} \label{events}
\end{figure}
At intermediate rigidities, $\Lambda_{TRD}$ and the velocity measured
with the RICH $\beta_{RICH}$ are used to separate the
antiproton signal from light particles ($e^-$ and $\pi^-$). As an
example, Fig. \ref{events}.a shows that the antiproton signal and the
background are well separated in the ($\beta_{RICH}$ - $\Lambda_{TRD}$) plane
for the absolute rigidity range 5.4-6.5 GV. To determine the
number of antiproton signal events, the $\pi^-$ background is
removed by a rigidity dependent $\beta_{RICH}$ cut and the $\Lambda_{TRD}$
distribution is used to construct the templates and to
differentiate between the $\bar{p}$ signal and $e^-$ background.
The background template is defined from the $e^-$ data
sample selected using ECAL. The Monte Carlo simulation
matches the data for $e^-$ events inside the ECAL. The
Monte Carlo simulation was then used to verify that the $e^-$
template shape outside the ECAL and inside the ECAL are
identical.
In the high rigidity region, the two-dimensional
($\Lambda_{TRD}$ - $\Lambda_{CC}$) distribution is used to determine the number
of antiproton signal events. The lower bound of $\Lambda_{CC}$ is
chosen for each bin to optimize the accuracy of the fit. For
example, for $\vert R \vert >$ 175 GV, $\Lambda_{CC}$ $\geq$ -0.6. Variation of the
lower bound is included in the systematic errors discussed
below. To fit the data three template shapes are defined. The
first two are for antiprotons and electrons with correctly
reconstructed charge sign and the last one is for charge
confused protons. The background templates (i.e., electrons
and charge confused protons) are from the
Monte Carlo simulation. The Monte Carlo simulation of
the charge confusion was verified with the 400 GV proton
test beam data. An example of the fit to the data is shown in
Figs. \ref{events}.b and \ref{events}.c for the rigidity bin 175-211 GV.
The distribution of data in the ($\Lambda_{TRD}$ - $\Lambda_{CC}$) plane is shown in
Fig. \ref{events}.b and the fit results showing the signal and background
distributions is highlighted in Fig. \ref{events}.c. The $\chi^2$ of
the fit is 138 for 154 degrees of freedom.
Overall, results for all 57 rigidity bins give a total of
3.49 $\times 10^5$ antiproton events in the data.

\noindent {\bf Analysis.}
The isotropic antiproton flux for the absolute
rigidity bin $R_i$ of width $\Delta R_i$ is given by
\begin{equation}\label{eq:phi}
\Phi_i^{\bar{p}}= \frac{N_i^{\bar{p}}}{A_i^{\bar{p}}T_i\Delta R_i}
\end{equation}
where the rigidity is defined on top of the AMS, $N^{\bar{p}}_i$
is the number of antiprotons in the rigidity bin $i$ corrected with
the rigidity resolution function \cite{ref:AMS2_pbar}. 
$A^{\bar{p}}_i$ is the corresponding effective acceptance
that includes geometric
acceptance as well as the trigger and selection efficiency,
and $T_i$ is the exposure time.
Detector resolution effects cause migration of events $N^{\bar{p}}_i$
from rigidity bin $R_i$ to the measured rigidity bins $\tilde{R}_j$
resulting in the observed number of events $\tilde{N}_i^{\bar{p}}$.
To account for this event migration, an iterative unfolding
procedure is used to correct the number of observed
events \cite{ref:AMS2_protons,ref:AMS2_pbar}.
The same procedure is used to unfold the observed number
of proton events.
The ($\bar{p}/p$) flux ratio is defined for each absolute rigidity
bin by:
\begin{equation}\label{eq:rat}
  \left(\frac{\bar{p}}{p}\right)_i = \frac{\Phi_i^{\bar{p}}}{\Phi_i^{p}}=\frac{\tilde{N}_i^{\bar{p}}}{\tilde{N}_i^{p}}\frac{\tilde{A}_i^{p}}{\tilde{A}_i^{\bar{p}}}
\end{equation}
where $\tilde{A}_i^{p}/\tilde{A}_i^{\bar{p}}$ is the ratio of folded acceptances.
We note that the $\tilde{A}_i^{p}/\tilde{A}_i^{\bar{p}}$
ratio decreases from 1.15 at 1 GV to 1.04 at 450 GV due to the varying difference of
interaction cross sections for protons and antiprotons (and
 considering bin-to-bin event migration).
With 3.49 $\times 10^5$ antiproton events, the accurate study of
systematic errors is the key part of the present analysis,
a detailed description can be found in \cite{ref:AMS2_pbar}.
Overall systematic error on the antiproton-to-proton flux ratio ranges
from $\sim 8\%$ at 1 GV to $\sim 13\%$ in the last bin (259-450 GV)
with a minimum of $\sim 2\%$ in the intermediate rigidity range ($\sim 30$ GV)
\cite{ref:AMS2_pbar}.

\begin{figure*}[t]
\centering
\includegraphics[width=133mm]{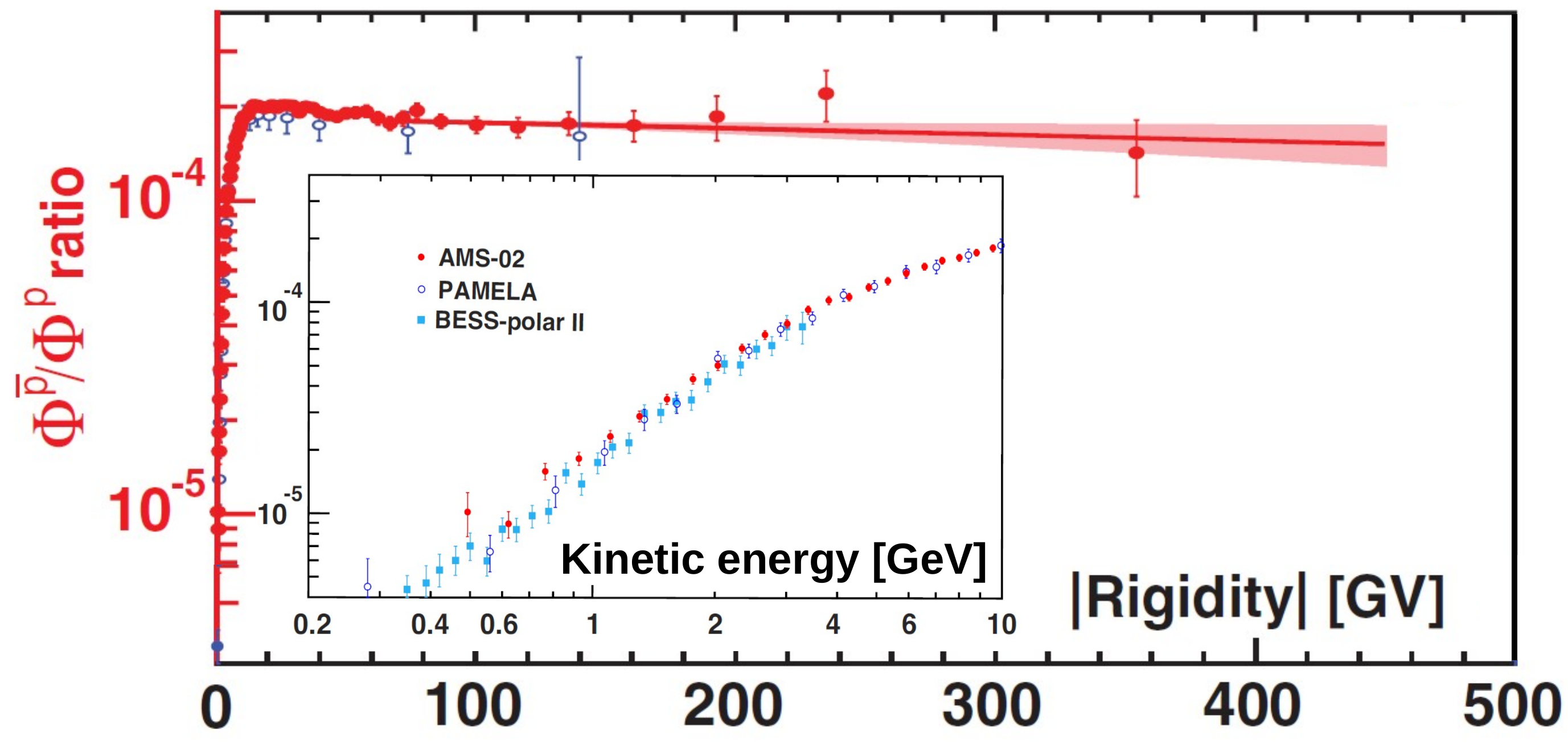}
\caption{The measured ($\bar{p}/p$) flux ratio as a function of the
  absolute value of the rigidity from 1 to 450 GV. The PAMELA \cite{ref:PAMELA}
  measurement is also shown (blue open circle).
  In the inset the ($\bar{p}/p$) flux ratio it is shown as a function of the kinetic energy up to 10 GeV.
  The kinetic energy is defined as $E_K = \sqrt{R^2 +M^2}-M$ where M is the antiproton or proton mass.
  The BESS \cite{ref:BESS} and PAMELA \cite{ref:PAMELA} measurements are also shown. For the AMS data, the error bars are
  the quadratic sum of statistical and systematic errors. Horizontally, the data points are placed at
the center of each bin. See \cite{ref:AMS2_pbar}.} \label{Ratio}
\end{figure*}

\noindent {\bf Results.}
The measured antiproton-to-proton flux ratio
as a function of the absolute rigidity value at the top of the AMS
is shown in Fig. \ref{Ratio}. 
The AMS results,
compared with earlier experiments,
extend the rigidity range to 450 GV with increased
precision.
The inset Fig. \ref{Ratio} shows the low energy ($<$ 10 GeV) part of the measured flux ratio.
The measured values of ($\bar{p}/p$) flux ratio, together with the statistical and systematic
errors can be found in Table I of Supplemental Material of \cite{ref:AMS2_pbar} and
is stored online in the ASI/ASDC cosmic ray database \cite{ref:ASDC} as well as all
the other published results from the AMS experiment.
The statistical errors are obtained from the fit errors on the signal, and
both statistical and systematic error contributions to the total error in the
flux ratio vary with rigidity.
For 1.00 $\leq \vert R \vert <$ 1.33 GV the statistical error
dominates, for 1.33 $\leq \vert R \vert <$ 1.71 GV the errors are comparable,
for 1.71 $\leq \vert R \vert <$ 48.5 GV the systematic error dominates,
for 48.5 $\leq \vert R \vert <$ 108 GV the errors are comparable,
and for 108 $\leq \vert R \vert <$ 450 GV statistical error dominates.
To minimize the systematic error for this flux ratio we have used the 2.42 $\times 10^9$
protons selected with the same acceptance, time period, and absolute rigidity range
as the antiprotons. From 10 to 450 GV, the values of the
proton flux are identical to 1$\%$ to those in our publication
\cite{ref:AMS2_protons}.
As seen from Fig. \ref{Ratio} the ($\bar{p}/p$) flux ratio reaches a maximum at $\sim 20$ GV and above $\sim$60  GV appears to be rigidity independent.
To estimate the lowest rigidity above which the ($\bar{p}/p$)
ratio is rigidity independent, we use rigidity intervals
with starting rigidities from 10 GV and increasing bin by
bin. The ending rigidity for all intervals is fixed at 450 GV.
Each interval is split into two sections with a boundary
between the starting rigidity and 450 GV. Each of the two
sections is fit with a constant and we obtain two mean
values of the ($\bar{p}/p$) ratio. The lowest starting rigidity of
the interval that gives consistent mean values at the
90$\%$ C.L. for any boundary defines the lowest limit.
This yields 60.3 GV as the lowest rigidity above which
the ($\bar{p}/p$) flux ratio is rigidity independent with a mean
value of 1.81 $\pm$ 0.04 $\times 10^{-4}$.
Further tests about the flatness of the measured ratio above $\sim 60$ GV
are described in \cite{ref:AMS2_pbar}.
It is interesting to note that traditional models 
for the secondary antiproton flux
are predicting a decreasing behaviour for the ($\bar{p}/p$) flux ratio
(see e.g \cite{ref:donato}).

\begin{acknowledgments}
  This work has been supported by acknowledged persons and institutions in the published papers about the AMS-02
  measurements \cite{ref:AMS2_first,ref:AMS2_fraction,ref:AMS2_epfluxes,ref:AMS2_allele,ref:AMS2_protons,ref:AMS2_He,ref:AMS2_pbar,ref:AMS2_BC}
  as well as by the Italian Space Agency under contracts ASI-INFN: C/011/11/1 - I/002/13/0 and I/037/14/0.
\end{acknowledgments}


\end{document}